\newif\ifemulateapj
\emulateapjtrue

\ifemulateapj
\documentclass[iop]{emulateapj}

\newcommand{\eqnewline}[1]{\nonumber \\ #1}

\else
\documentclass[12pt,preprint]{aastex}
\usepackage{lineno}
\linenumbers

\newcommand{\eqnewline}[1]{}

\fi

\usepackage{amsmath,amssymb}

\newcommand{\del}{\partial}
\newcommand{\mb}[1]{\mathbf{#1}}
\newcommand{\rot}{\mathbf{\nabla} \times}
\renewcommand{\div}{\mathbf{\nabla} \cdot}
\newcommand{\G}{\Gamma}

\newcommand{\FigureOne}{
\begin{figure}[tb]
 \figurenum{1}
 \ifemulateapj
 \plotone{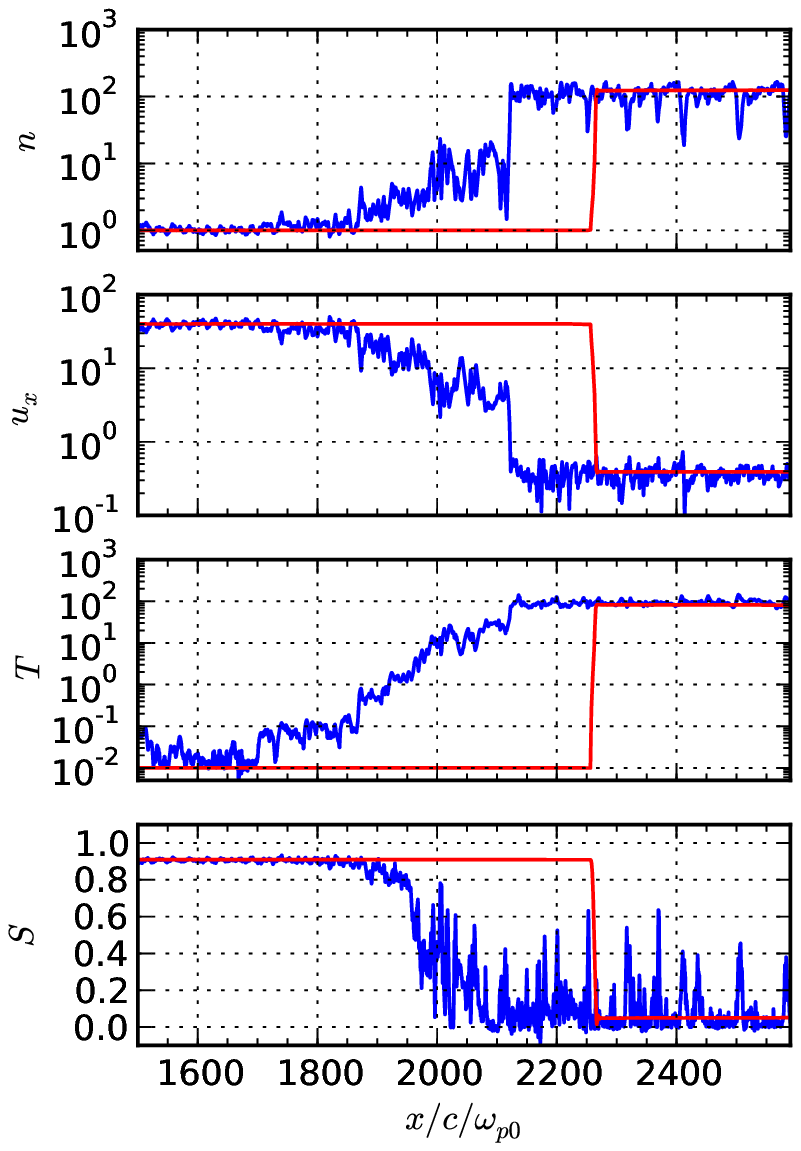}
 \else
 \includegraphics[scale=1.0]{figure1}
 \fi

 \caption{Comparison of the shock structures in run~A (blue) and B (red) at
 $\omega_{p0} t = 1700$. From top to bottom, each panel shows spatial profiles
 of proper density, longitudinal component of four velocity, temperature, and
 Poynting flux.} \label{fig1}
\end{figure}
}

\newcommand{\FigureTwo}{
\begin{figure}[tb]
 \figurenum{2}
 \ifemulateapj
 \plotone{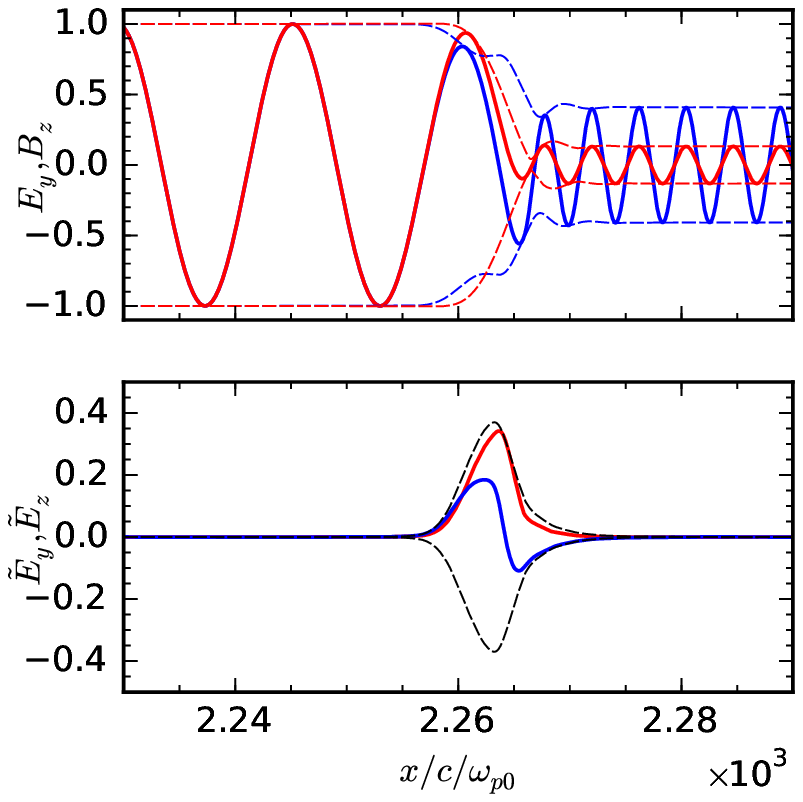}
 \else
 \includegraphics[scale=1.0]{figure2}
 \fi

 \caption{Enlarged view of electromagnetic fields around the shock-like
 discontinuity at $\omega_{p0} t = 1700$ in run~B. The top panel shows raw
 electromagnetic fields $E_y$ (solid red line), $B_z$ (solid blue line), $\pm
 \sqrt{E_y^2 + E_z^2}$ (dashed red line), and $\pm \sqrt{B_y^2 + B_z^2}$
 (dashed blue line) whereas the bottom one shows non-MHD electric fields
 $\tilde{E}_y$ (red line) and $\tilde{E}_z$ (blue line), as well as the
 absolute value $\pm \sqrt{\tilde{E}_y^2 + \tilde{E}_z^2}$ (dashed black
 line).}  \label{fig2}
\end{figure}
}

\newcommand{\FigureThree}{
\begin{figure}[tb]
 \figurenum{3}
 \ifemulateapj
 \plotone{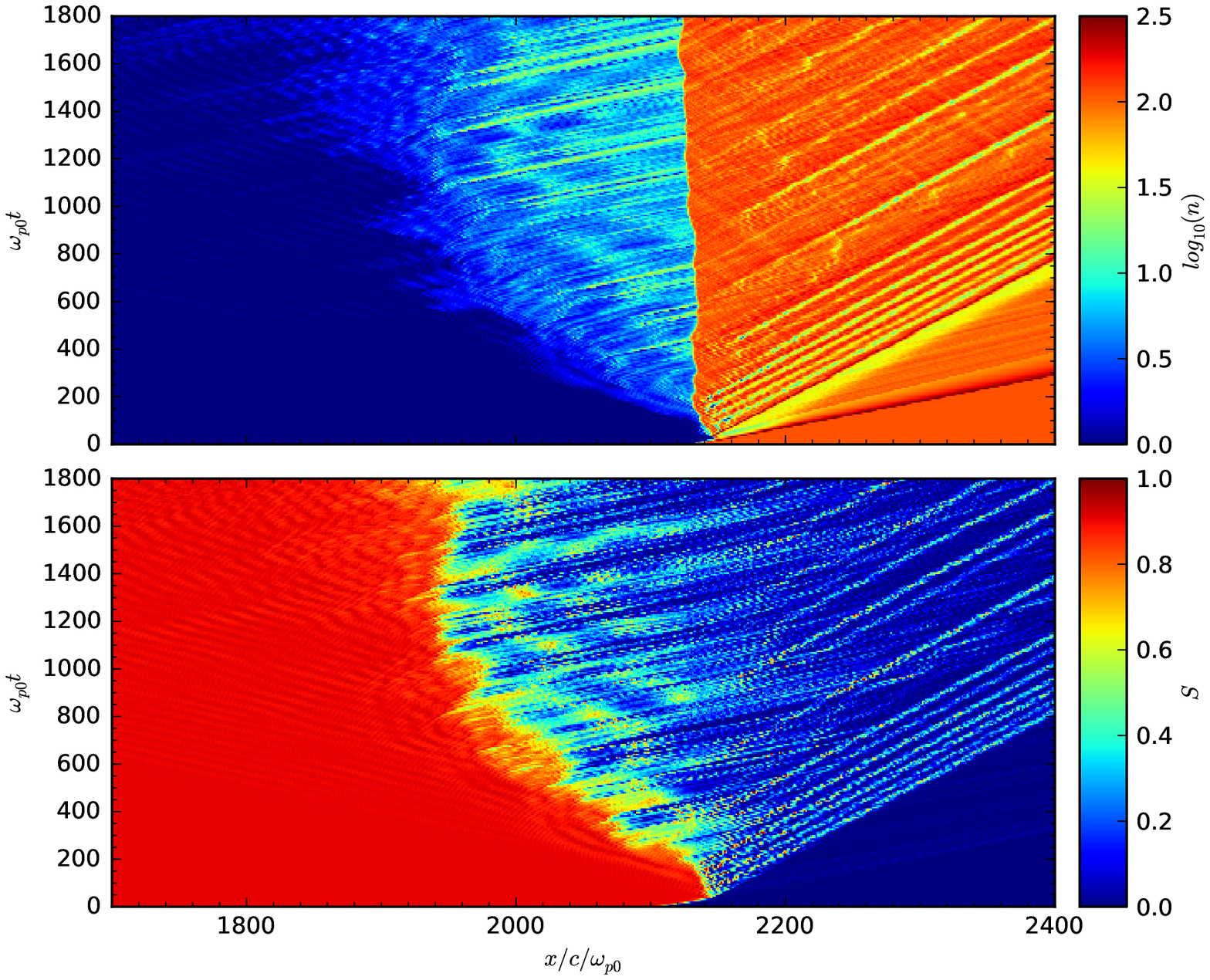}
 \else
 \includegraphics[scale=0.8]{figure3}
 \fi

 \caption{Time evolution of the modified shock structure in run~A. The top and
 bottom panels respectively show proper density and Poynting flux.}
 \label{fig3}
\end{figure}
}

\newcommand{\FigureFour}{
\begin{figure*}[tb]
 \figurenum{4}
 \ifemulateapj
 \plotone{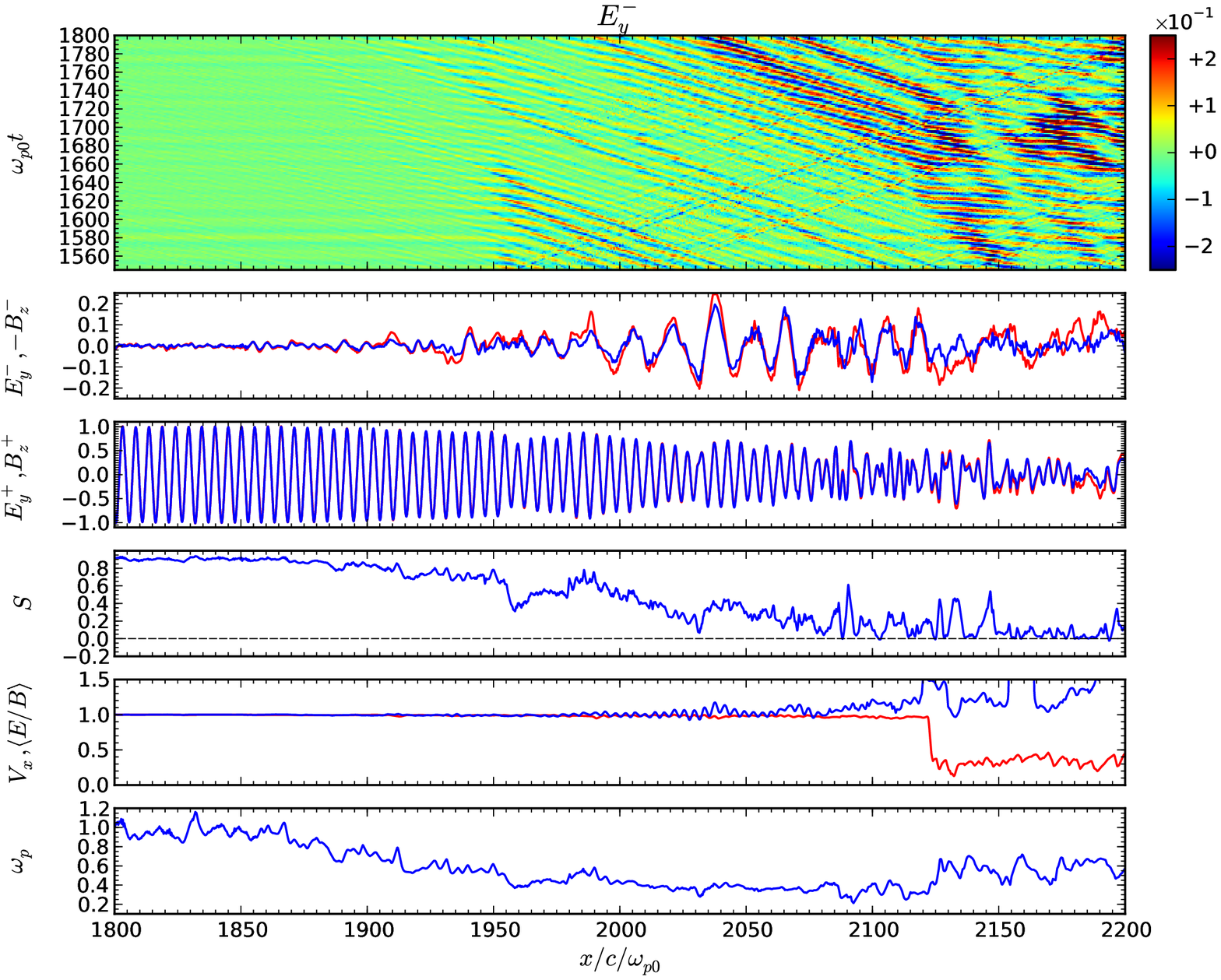}
 \else
 \includegraphics[scale=0.8]{figure4}
 \fi

 \caption{Enlarged view of the precursor region at $\omega_{p0} t = 1800$ in
 run~A. The top panel shows time evolution of $E_y^{-}$ (negative helicity) in
 the precursor. Moving downwards, other panels represent the negative helicity
 components $E_y^{-}$ (red) and $-B_z^{-}$ (blue), the positive helicity
 components $E_y^{+}$ (red) and $B_z^{+}$ (blue), the Poynting flux,
 three-velocity $V_x$ (red) and the ratio $\left<E/B\right>$ (blue), and the
 proper plasma frequency $\omega_p$ in the bottom, respectively. Note that $\left<E/B\right>$ is a moving average of the raw profile over a spatial
 window of width $10 c/\omega_{p0}$.} \label{fig4}
\end{figure*}
}

\newcommand{\FigureFive}{
\begin{figure*}[tb]
 \figurenum{5}
 \ifemulateapj
 \plotone{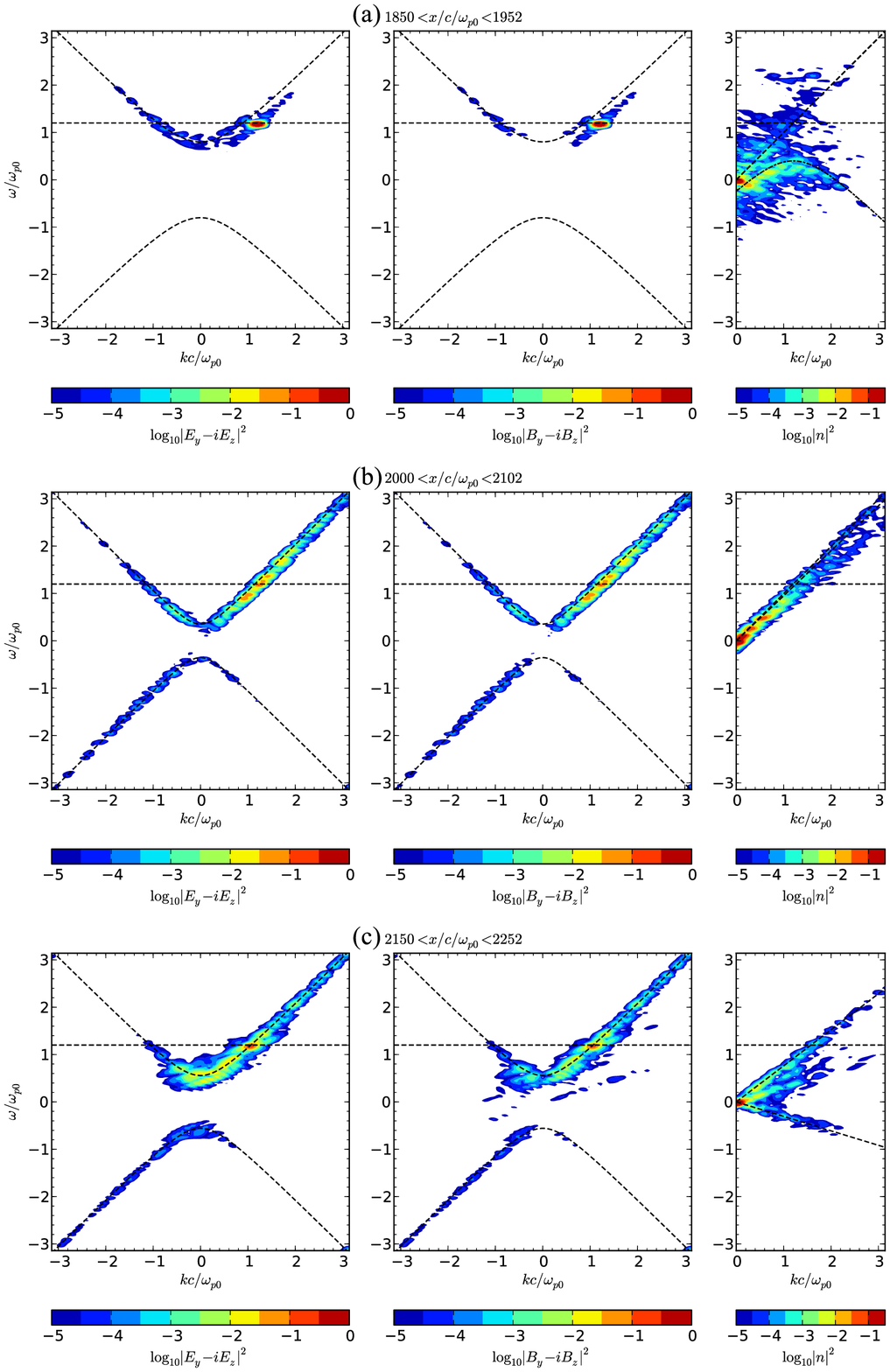}
 \else
 \includegraphics[scale=0.75]{figure5}
 \fi

 \caption{Power spectra of (left) $E_y - i E_z$, (center) $B_y - i B_z$,
(right) $n$ in $k$~-~$\omega$ space obtained from run~A. The spectra are
calculated for the time interval $1544 \leq \omega_{p0} t \leq 1800$ and
spatial intervals corresponding to (a) the precursor edge $1850 <
x/c/\omega_{p0} < 1952$, (b) deep in the precursor $2000 < x/c/\omega_{p} <
2102$, and (c) downstream $2150 < x/c/\omega_{p} < 2252$.}  \label{fig5}
\end{figure*}
}

\newcommand{\FigureSix}{
\begin{figure*}[tb]
 \figurenum{6}
 \ifemulateapj
 \plotone{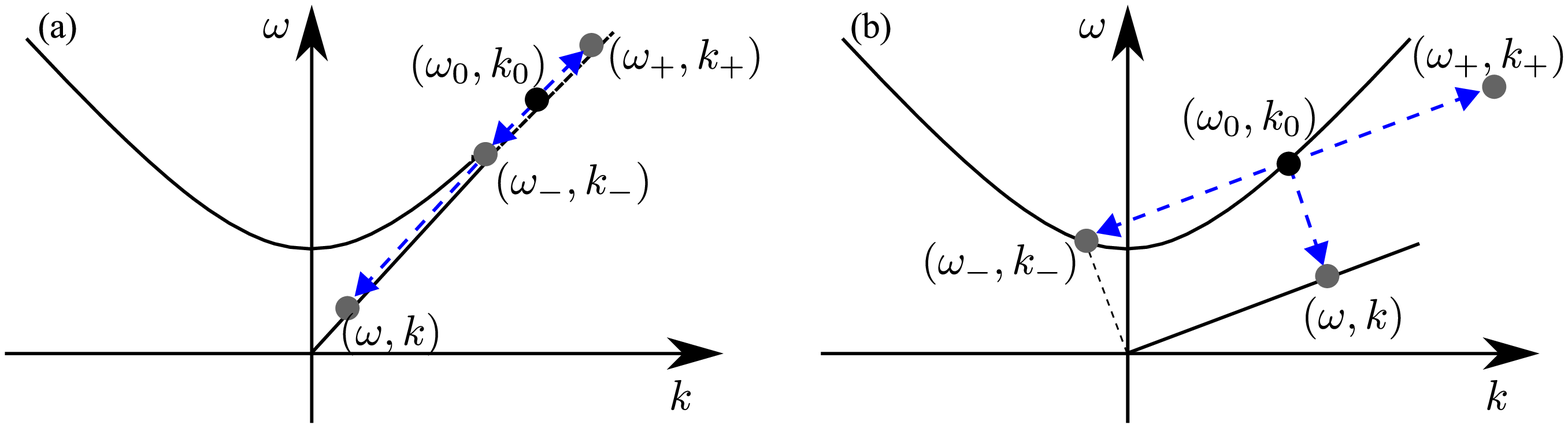}
 \else
 \includegraphics[scale=0.6]{figure6}
 \fi

 \caption{Schematic illustration of parametric instabilities likely to occur
 in (a) precursor, (b) downstream. The upper curve represents the nonlinear
 dispersion relation of a superluminal wave. The lower straight line indicates
 the sound-mode dispersion relation. Note that the forward and backward sound
 waves in (a) almost coincide with $\omega = k c$ and are indistinguishable
 because the plasma flows with relativistic bulk speed in the laboratory
 frame.} \label{fig6}
\end{figure*}
}

\newcommand{\FigureSeven}{
\begin{figure}[tb]
 \figurenum{7}
 \ifemulateapj
 \plotone{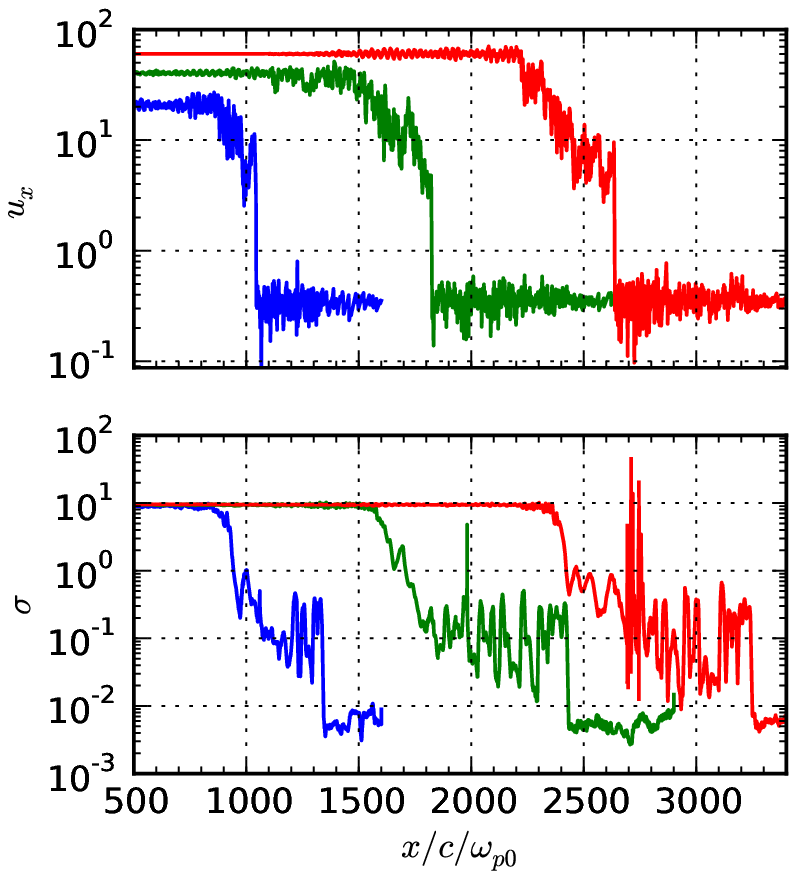}
 \else
 \includegraphics[scale=0.8]{figure7}
 \fi

 \caption{The dependence on the upstream Lorentz factor $\gamma_0$ of the flow
 four-velocity $u_x$ (top) and the magnetization $\sigma$ (bottom). Results
 are shown for $\gamma_0 = 20$ (blue), $\gamma_0 = 40$ (green), $\gamma_0 =
 60$ (red). In each case $\sigma_0 = 10$ and $\Omega/\omega_{p0} = 1.2$. The
 curves are shifted in the horizontal direction for clarity.} \label{fig7}
\end{figure}
}

\shorttitle{Role of Superluminal Waves in Termination Shocks}
\shortauthors{Amano \& Kirk}

\begin{document}

\title{The Role of Superluminal Electromagnetic Waves in Pulsar Wind Termination Shocks}
\author{Takanobu Amano}
\email{amano@eps.s.u-tokyo.ac.jp}
\affil{Department of Earth \& Planetary Science, University of Tokyo, 113-0033, Japan}
\author{John G. Kirk}
\affil{Max-Planck-Institut f\"ur Kernphysik, Postfach 103980, 69029 Heidelberg, Germany}

\begin{abstract}
 The dynamics of a standing shock front in a Poynting-flux dominated
 relativistic flow is investigated by using a one-dimensional, relativistic,
 two-fluid simulation. An upstream flow containing a circularly polarized,
 sinusoidal magnetic shear wave is considered, mimicking a wave driven by an
 obliquely rotating pulsar. It is demonstrated that this wave is converted
 into large amplitude electromagnetic waves with superluminal phase speeds by
 interacting with the shock when the shock-frame frequency of the wave exceeds
 the proper plasma frequency. The superluminal waves propagate in the
 upstream, modify the shock structure substantially, and form a well-developed
 precursor region ahead of a subshock. Dissipation of Poynting flux occurs in
 the precursor as well as in the downstream region through a parametric
 instability driven by the superluminal waves. The Poynting flux remaining in
 the downstream region is carried entirely by the superluminal waves. The
 downstream plasma is therefore an essentially unmagnetized, relativistically
 hot plasma with a non-relativistic flow speed, as suggested by observations
 of pulsar wind nebulae.
\end{abstract}

\keywords{plasmas --- pulsars: general --- stars: winds, outflows --- waves}

\section{Introduction}

Poynting-flux dominated, relativistic outflows are thought to emerge from
compact objects in several high energy astrophysical environments, including
pulsar winds, jets in active galactic nuclei and gamma-ray bursts. A key issue
in this context is how the energetically dominant Poynting flux is converted
into particles, and, subsequently, into the observed radiation; this problem,
known as the $\sigma$-problem, was originally formulated for the Crab pulsar
wind \citep{1974MNRAS.167....1R,1984ApJ...283..694K,1998MmSAI..69.1009M}, but
is now recognized in a more generic context
\citep[e.g.,][]{1994MNRAS.270..480T,2002A&A...391.1141D}.

A rotation-powered pulsar presumably emits most of its spin-down luminosity in
the form of a relativistic wind that powers the surrounding pulsar wind nebula
(PWN), in which relativistic electrons and positrons emit synchrotron
radiation. These pairs are believed to be produced by a cascade that occurs
close to a rotating, strongly magnetized neutron star, where an enormous
electric field is induced
\citep{1971ApJ...164..529S,1975ApJ...196...51R,2001ApJ...560..871H}. Although
the efficiency of pair production is rather uncertain, all models predict a
Poynting-flux dominated wind, i.e., the parameter $\sigma$, defined as the
ratio of the Poynting flux to the particle kinetic energy flux, is believed to
be much greater than unity $\sigma \gg 1$. In a supersonic, radially expanding
ideal MHD wind, $\sigma$ is constant, so that the wind remains Poynting flux
dominated until it reaches the termination shock. Beyond the termination
shock, however, the most of the energy flux seems to be carried by particles
($\sigma \ll 1$) according to arguments based on the morphology of the PWN and
the spectrum of its radiation
\citep{1974MNRAS.167....1R,1984ApJ...283..694K}. This apparent contradiction
has been the subject of considerable debate over the years. It suggests the
existence of an unknown dissipation mechanism that converts the dominant
electromagnetic energy into particle kinetic energy inside the wind zone with
substantial efficiency, although the constraint might be partially relaxed by
an MHD instability in the inner part of the PWN
\citep{1998ApJ...493..291B,2011ApJ...728...90M,2013MNRAS.tmpL..60P}.

One approach to resolve the problem is to invoke an oblique rotator, whose
rotation axis is not parallel to the magnetic axis
\citep{1990ApJ...349..538C,1994ApJ...431..397M}. Such an obliquely rotating
pulsar will continuously launch a wave along with the wind because of
time-varying electromagnetic fields of the magnetosphere. The plasma
populating the magnetosphere is dense enough to screen out the electric fields
of such a wave, which, therefore, propagates essentially as an MHD wave. Being
an MHD wave, it is conceivable that only a quasi-static structure survives in
the wind zone far from the star. The striped wind model, first proposed by
\cite{1990ApJ...349..538C}, is based on such an idea. In such a wind, toroidal
magnetic fields of alternating polarity, separated by a neutral sheet, are
embedded in the flow. The particles inside the neutral sheet provide the
pressure to balance the magnetic pressure outside the sheet. As the wind
expands radially, the plasma pressure tends to fall off adiabatically and more
rapidly than the magnetic pressure, resulting in the compression of the
sheet. There is obviously a minimum thickness of the sheet, below which the
MHD approximation is violated. A reasonable estimate of this is the Larmor
radius of the pairs defined with the temperature inside and the magnetic field
outside the sheet. If the sheet were to be compressed beyond this limit,
non-ideal MHD effects would start to play a role. It has been suggested that
magnetic reconnection may occur because of enhanced anomalous resistivity in
such a thin neutral sheet, an idea that is supported by Particle-In-Cell (PIC)
simulations
\citep{2001ApJ...562L..63Z,2005PhRvL..95i5001Z,2007ApJ...670..702Z,2004PhPl...11.1151J}.
In order to maintain pressure balance, the plasma inside the sheet is heated
by annihilating magnetic energy. As the flow propagates farther out, this
process may proceed in a quasi-static manner until the stripe structure is
completely erased. However, the heating of the plasma by dissipation exerts a
pressure gradient on the plasma, which leads to substantial acceleration of
the flow \citep{2001ApJ...547..437L,2003ApJ...591..366K}, and the apparent
dissipation rate decreases because of the relativistic time dilation
effect. Consequently, \citet{2001ApJ...547..437L} concluded that the flow
remains Poynting flux dominated until it reaches the termination shock.

The failure of damping mechanisms to dissipate the wave before it reaches the
location of the termination shock has motivated a more detailed study of the
interaction between current sheets and a relativistic shock
\citep{2003MNRAS.345..153L,2007A&A...473..683P,2008ApJ...680..627N,2011ApJ...741...39S}.
\citet{2003MNRAS.345..153L} suggested that the dissipation of Poynting flux
may occur due to driven magnetic reconnection triggered at current sheets
encountered by the termination shock. By conducting two- and three-dimensional
PIC simulations, \cite{2011ApJ...741...39S} have recently confirmed that
magnetic reconnection is indeed driven by the interaction with the shock, and
that the resulting distribution in energy of the accelerated particles agrees
with that observed in PWNe. However, the pair production rate assumed in these
simulations is much larger than suggested previously
\citep{2012SSRv..tmp...33A}, and would, according to the estimates of
\citet{2003ApJ...591..366K}, result in substantial damping of the striped wind
before it encounters the shock.

The above picture is based on an MHD description; although non-ideal MHD
effects cause the dissipation, the structure (or wave) itself can be described
in the language of MHD with appropriate transport coefficients. Since the
frozen-in condition requires that the electric field is smaller than the
magnetic field $E < B$, the phase velocities of MHD waves are always
subluminal ($E/B$ gives a phase velocity). But when one moves beyond the MHD
model, additional degrees of freedom open up that allow a wave with $E > B$ to
propagate. Although the wave amplitude, under pulsar conditions, is strongly
nonlinear, such a wave is essentially an electromagnetic wave in vacuum
modified by the presence of a plasma
\citep[e.g.,][]{Akhiezer1956,1971PhRvL..27.1342M}, and has long been studied
in the context of pulsar physics
\citep[e.g.,][]{1976JPlPh..15..335K,1996MNRAS.279.1168M,2010PPCF...52l4029K,2012ApJ...745..108A}.
We will call this electromagnetic wave {\em superluminal} because its phase
speed distinguishes it from subluminal MHD waves (which also involve
electromagnetic fluctuations). An obvious advantage of superluminal modes is
their expected large damping rates through instabilities
\citep[e.g.,][]{1973PhFl...16.1480M,1975A&A....41..431S,1978A&A....65..401A,1978JPlPh..20..313L},
which suggests they might efficiently dissipate either before or at the
termination shock \citep{2005ApJ...634..542S}.

There exists a cutoff frequency below which a superluminal wave cannot
propagate due to screening of the wave electric field by the plasma particles,
suggesting that these modes can play a role only at sufficiently large
distance from a pulsar. More specifically, nonlinear solutions for
superluminal waves have been found both for circularly and linearly polarized
modes, and used to construct and solve jump condition between a subluminal (or
striped) mode and superluminal modes that carry the same particle, momentum,
and energy fluxes \citep{2010PPCF...52l4029K,2012ApJ...745..108A}. These
studies formulate necessary conditions for the conversion to a superluminal
wave to occur, and found it to be possible beyond a critical radius, which is
located well inside the termination shock in the equatorial region for an
isolated pulsar. However, they did not address the question of how and where
the conversion actually takes place. This could occur spontaneously when the
magnetic reconnection proceeds too slowly to maintain the pressure balance
required by the radial evolution of the wind. Alternatively, it may be
triggered by the interaction with the termination shock. It is this latter
possibility which we study here.

Using a newly developed, one-dimensional, relativistic, two-fluid simulation
code for pair plasmas, we investigate the roles played by superluminal waves
in a relativistic shock. A circularly polarized sinusoidal magnetic shear wave
is adopted to mimic a pulsar-driven wave, and is forced to interact with a
termination shock. We demonstrate that superluminal waves are indeed
generated, strongly modify, and, eventually, dominate the shock structure. A
well developed precursor region is formed ahead of a subshock, in which the
flow is decelerated and substantial plasma heating occurs due to efficient
dissipation of Poynting flux. We argue that the basic process that leads to
the dissipation and modification of the shock is a parametric
instability: the scattering of superluminal waves off a longitudinal
density perturbation causes the energy and momentum of the wave to be
transferred to the longitudinal mode. When these perturbations grow to large
amplitudes, they steepen to form small-scale shocks that eventually cause the
required dissipation.

In Section~\ref{model}, the basic equations and the simulation setup used in
this study are presented. The results are discussed in Section~\ref{result},
in which a detailed analysis of the shock structure, as modified by
superluminal waves, is given. Section~\ref{discussion} contains a discussion
of the astrophysical application and points out some remaining open issues,
and Section~\ref{summary} presents a summary.

\section{Model}
\label{model}

\subsection{Basic Equations}

The simplest possible model that is capable of describing the superluminal
electromagnetic waves on which we focus in our study is that of two
relativistic fluids (electrons and positrons), supplemented by Maxwell's
equations. Unlike most previous studies, which considered cold fluids, we
include a finite pressure for each species, that is determined by a polytropic
equation of state. This leads to the following set of equations:
\begin{eqnarray}
 && \frac{\del}{\del t} \left( \gamma_s n_s \right) +
  \div \left( n_s \mb{u}_s \right) = 0, \\
 && \frac{\del}{\del t} \left( \frac{w_s}{c^2} \gamma_s \mb{u}_s \right) +
  \div \left( \frac{w_s}{c^2} \mb{u}_s \mb{u}_s + \mb{I} p_s \right) =
  \eqnewline{&& \qquad}
  q_s \gamma_s n_s
  \left(\mb{E} + \frac{\mb{u}_s}{\gamma_s c} \times \mb{B}\right), \\
 && \frac{\del}{\del t} \left( w_s \gamma_s^2 - p_s \right) +
  \div \left( w_s \gamma_s \mb{u}_s \right) =
  q_s n_s \mb{u}_s \cdot \mb{E}, \\
 && \frac{1}{c} \frac{\del}{\del t} \mb{E} = \rot \mb{B} +
  \frac{4 \pi}{c} \mb{J}, \\
 && \frac{1}{c} \frac{\del}{\del t} \mb{B} = -\rot \mb{E}, \\
 && \div \mb{E} = 4 \pi \rho, \label{coulomb} \\
 && \div \mb{B} = 0, \label{divb}
\end{eqnarray}
where $n_s$, $(\gamma_s c, \mb{u}_s)$, $p_s$ are the proper number density,
four velocity, and proper pressure of particle species $s$, respectively. The
enthalpy density $w_s = n_s m_s c^2 + \G/(\G-1) p_s$ is written in terms of
the ratio of specific heat $\G$ ($\G = 4/3$ is used throughout in this
study). The charge density $\rho = \gamma_{p}q_{p} n_{p} + \gamma_{e}q_{e}
n_{e}$ and current density $\mb{J} = q_{p} n_{p} \mb{u}_{p} + q_{e} n_{e}
\mb{u}_{e}$ introduce the coupling between fluids and electromagnetic
fields. Notice that the subscripts ($p$ for positrons, $e$ for electrons) for
particle species are omitted for brevity whenever no confusion arises. Other
notations are standard: the speed of light $c$, electron mass $m$, elementary
charge $e$ (thus $q_{p} = - q_{e} = e$). In the current one-dimensional
simulation, equations~(\ref{coulomb}) and (\ref{divb}) are automatically
satisfied throughout the simulation, provided they are satisfied by the
initial and boundary conditions. Note that we solve the fluid equations
without any assumption on symmetry or antisymmetry so that Langmuir waves
arising from oscillations in charge density are included in the
model. However, the initial conditions are such that the longitudinal
(transverse) components of the fluid velocities are symmetric (antisymmetric)
and there are no charge density perturbations. There is, therefore, no
Langmuir wave activity initially. In fact, Langmuir waves remain unimportant
throughout the simulations, for reasons we discuss in
section~\ref{discussion}. This allows us to concentrate on the dynamics of the
positron fluid in our discussion.

One effect introduced by the finite temperature, $T=p/n$, is the effective
increase of particle inertia by a factor of $h = w / (n m c^2) = 1 + \G/(\G-1)
T/m c^2$.  This has an important consequence when considering superluminal
waves. In Appendix \ref{superluminal}, we present an exact solution of a
circularly polarized superluminal wave whose cutoff frequency is determined by
an effective proper plasma frequency
\begin{eqnarray}
 \omega_{p} \equiv \sqrt{\frac{8 \pi n e^2}{m h}} \label{eq:cutoff}
\end{eqnarray}
with $n = n_{p} = n_{e}$. Thus, the cutoff frequency of a superluminal wave is
reduced, because of the increased effective inertia in a relativistically hot
plasma $T \gg mc^2$.

In our numerical implementation, a different, but mathematically equivalent,
set of equations is used, which enables us to solve the time evolution without
violating important conservation laws. A detailed description is presented in
Appendix~\ref{numerical}.

\subsection{Simulation Setup}

We are interested in quasi steady-state solutions in which a relativistic,
strongly magnetized flow dissipates its Poynting flux. Therefore, we attempt
to initialize the system so that it will quickly reach such a solution, if it
exists.  We divide the one-dimensional simulation box initially into two
regions. The left-hand (upstream) side contains a cold, highly magnetized
plasma flowing to the right with supersonic speed, and the right-hand
(downstream) side contains an unmagnetized, relativistically hot plasma.

The upstream plasma carries only an oscillating component of the magnetic
field, i.e., its phase-averaged value is zero. Such a situation is believed to
be realized at the equator in the striped wind model. Although the wave in the
striped wind is likely to be a linearly polarized entropy mode, containing hot
current sheets, we adopt here a much simpler model. We assume the upstream
wave is a circularly polarized sinusoidal magnetic shear. In the comoving
frame of the upstream fluid, it is a stationary structure, in which the
magnetic field vector rotates in space without changing its absolute
value. The density and pressure are, accordingly, constant. Since the phase
speed of this wave vanishes in the co-moving frame, it remains subluminal in
all physically relevant frames of reference. The wave is a limiting case of
the subluminal mode discussed by \citet{2002ASPC..271..115M}, taken as the
frequency and the phase-independent component of the magnetic field tend to
zero (in his notation $\eta\rightarrow\infty$); its radial evolution has been
re-examined recently in the context of blazar jets
\citep{2011ApJ...729..104K,2011ApJ...736..165K}. In the context of laboratory
plasma physics, this mode is known as a \lq\lq sheet-pinch\rq\rq\
\citep{2003PhPl...10.2763L} configuration. We employ this substantial
simplification because we believe that the most important parameter in the
problem is the ratio of the wave frequency $\Omega$ to the proper plasma
frequency $\omega_{p}$, a view which is supported by the PIC simulations of
\citet{2011ApJ...741...39S}. This implies that the details of the wave mode,
such as polarization and functional form, will not have an important
qualitative influence on the results. One advantage of using circular
polarization is that it greatly simplifies the theoretical analysis.

In the following, the upstream proper positron density, bulk flow Lorentz
factor in the simulation frame, and proper temperature are
denoted respectively by $n_0$, $\gamma_0$, and $T_0$. The upstream
magnetization parameter is defined as $\sigma_0 = B_0^2 / 8 \pi
\gamma_0^2 n_0 m c^2$, where $B_0$ is the magnetic field strength in
the upstream. To measure the dissipation in simulation results, we use
a more accurate and general definition of the magnetization given by
$\sigma = c (\mb{E} \times \mb{B})_x / (4 \pi \sum_{s} \gamma_s w_s
u_{s,x})$, which includes the finite temperature correction and does
not assume any relationship between $\mb{E}$ and $\mb{B}$.

The functional form of the upstream magnetic field in the simulation frame can
be given as
\begin{eqnarray}
 B_y &=& + B_0 \cos \left(k_0 x - \Omega t\right) \\
 B_z &=& - B_0 \sin \left(k_0 x - \Omega t\right),
\end{eqnarray}
where $\Omega > 0$ is the frequency and the wavenumber is $k_0 = \Omega / V_0
> 0$, with $V_0 = \sqrt{1 - 1/\gamma_0^2}$ being the upstream three-velocity.
The electric field is written according to the frozen-in condition $\mb{E} = -
V_0 \mb{e}_x \times \mb{B} / c$ ($\mb{e}_x$ is the unit vector in the $x$
direction). The transverse four-velocity is determined by Ampere's law to be
\begin{eqnarray}
 u_y &=& + u_{\perp,0} \cos \left(k_0 x - \Omega t\right) \\
 u_z &=& - u_{\perp,0} \sin \left(k_0 x - \Omega t\right),
\end{eqnarray}
for positrons with $u_{\perp,0} = k_0 B_0 / 8 \pi n_0 e$. The density,
temperature and the bulk flow velocity are not affected by the presence of
this wave. This magnetic shear structure is an analytic equilibrium solution
to the two-fluid equations, which is simply convected by the flow in the far
upstream.

For the sake of convenience, we define our terminology for the polarization
and helicity of transverse waves, which basically follows standard plasma
physics conventions. Notice, however, that the standard definition cannot be
used in a strict sense because it defines the sense of rotation with respect
to the static magnetic field, which is absent in our case. Instead, we simply
define the sign with respect to the $x$ axis. The helicity describes the sense
of rotation of a fluctuating vector in the $y$~-~$z$ plane, as $x$ increases
at a fixed time; when viewed along the positive $x$-direction, it is
anti-clockwise for positive, and clockwise for negative helicity. Similarly,
right-hand (R) and left-hand (L) polarization indicate that the rotation of a
vector in time measured at fixed $x$ is clockwise and anti-clockwise,
respectively. With this definition, the injected magnetic shear wave is
right-hand polarized ($R$) and has positive helicity ($+$), which we denote by
$R^{+}$. Note that the sign of superscript does not indicate the wave
propagation direction: $R^{+}$ and $L^{-}$ waves propagate in the positive,
while $R^{-}$ and $L^{+}$ waves propagate in the negative $x$
direction. Defining a complex quantity for a vector field, e.g., $B_y - i
B_z$, it is easy to confirm that the first, second, third, and fourth
quadrants of its $k$~-~$\omega$ Fourier space correspond to $R^{+}, R^{-},
L^{-}, L^{+}$, respectively. Consequently, a given vector quantity (say $E_{y,
z}$) can be decomposed into positive and negative helicity components by
Fourier transforming the complex quantity $E_y - i E_z$ into $k$-space,
extracting only the $k > 0$ ($k < 0$) Fourier components, and inverse
transforming back into configuration space, to find the positive (negative)
helicity component \citep{1986JGR....91.4171T}. This decomposition is used
below in the analysis of the simulation results.

The initial conditions in the downstream (right-hand) side are found from the
Rankine-Hugoniot relations under the assumption that the Poynting flux has
completely dissipated. The conservation laws in this case give
\begin{eqnarray}
 && 2 n_1 u_{x, 1} = 2 n_2 u_{x, 2}
\label{massflux}\\
 && 2 w_1 \frac{u_{x,1}^2}{c^2} + 2 p_1 +
  \left(1 + \frac{u_{x,1}^2}{\gamma_1^2 c^2}\right)
  \frac{B_1^2}{8 \pi} = 2 w_2 \frac{u_{x,2}^2}{c^2} + 2 p_2 \eqnewline{}
  \label{momflux}\\
 && 2 w_1 \gamma_1 u_{x,1} + \frac{u_{x,1}}{\gamma_1} \frac{B_1^2}{4 \pi} =
  2 w_2 \gamma_2 u_{x,2},
\label{energyflux}
\end{eqnarray}
where subscript 1 and 2 denote quantities in the upstream and downstream
regions. We retain a (small) contribution from the shear wave to the upstream
Lorentz factor, so that $\gamma_1^2 = \gamma_0^2 (1 + u_{\perp,0}^2)$ and
$u_{x,1} = \sqrt{1 + u_{\perp,0}^2} \gamma_0 V_0$. The other upstream
quantities ($w_1$, $p_1$ and $B_1$) are the same as those with subscript
0. For a strong shock with an upstream Lorentz factor $\gamma_0 \gg 1$ in a
cold plasma, the downstream flow speed $u_2$ and temperature $T_2$ may be
approximated as
\begin{eqnarray}
 u_2 &\simeq& \frac{\Gamma-1}{\sqrt{\Gamma(2-\Gamma)}} \\
 T_2 &\simeq& (\Gamma-1)\sqrt{\frac{2-\Gamma}{\Gamma}} \, \gamma_0 (1 + \sigma_0)
\end{eqnarray}
in the shock rest frame. The proper density is determined by the mass flux
conservation. The formal difference from a strong shock in an unmagnetized
plasma is the factor $1 + \sigma_0$ in the downstream temperature. This
additional temperature increase comes from the dissipation of Poynting
flux. We set up the initial downstream state by numerically solving equations
(\ref{massflux}), (\ref{momflux}), and (\ref{energyflux}) without additional
approximation.

In between the upstream and downstream regions, there exists, at the initial
instant, a buffer region in which the upstream magnetic field smoothly
decreases to zero. Since the physical quantities in the upstream and
downstream regions are initialized to be in pressure balance, the resulting
``shock'' structure would remain stationary in the simulation frame if such a
quasi-steady solution really exists.

The upstream wave is continuously injected from the upstream boundary during
the whole run. We choose the location of the upstream boundary so that any
perturbations generated by the interaction (propagating with the speed of
light) do not reach the boundary. On the other hand, electromagnetic waves may
reach the downstream boundary. We assume $\partial/\partial x = 0$ at this
boundary. Thus, in principle, there would be small but finite reflection of
waves. As we will see below, however, Poynting flux of electromagnetic waves
reaching the boundary is found to be very small. Therefore, we believe that
the effect of boundary conditions does not affect our discussion.

The following normalizations are used throughout the discussion;
$\omega_{p0}^{-1}$ for time, $c/\omega_{p0}$ for space, $c$ for velocity,
$n_0$ for density, $B_0$ for electric and magnetic fields, $\mu_0 = \gamma_0^2
(1 + \sigma_0) n_0 m c^3$ for energy flux density, respectively. Here,
$\omega_{p0} = \sqrt{8 \pi n_0 e^2/m}$ is the proper plasma frequency in the
upstream but without the relativistic temperature correction. We use a
constant upstream temperature $T_0/m c^2 = 0.01$, so that the correction is
not important in the upstream. The grid size and time step are chosen to be
$\Delta x = 0.05 c/\omega_{p0}$ and $\Delta t = 0.005 / \omega_{p0}$, and are
kept constant.

\section{Simulation Results}
\label{result}

\subsection{Overview}
\label{frequency}

First of all, we contrast two simulation results, one with a high, the other
with a low frequency $\Omega$, in Fig.~\ref{fig1}. We call the high-frequency
simulation with $\Omega/\omega_{p0} = 1.2$ ``run~A'', and the low frequency
simulation with $\Omega/\omega_{p0} = 0.4$ ``run~B''.  In each case the
upstream Lorentz factor is $\gamma_0 = 40$ and the magnetization is $\sigma_0
= 10$. One immediately sees that the shock structures are completely
different.

\ifemulateapj
\FigureOne
\fi

Run~A exhibits fluctuations of substantial amplitude in all the fluid
quantities, whereas the structure of run~B is uniform except for a
discontinuity, very much like a solution typically found in standard fluid
simulations. Although a discontinuity is also present, run~A shows a precursor
region characterized by a substantial deceleration of the flow and an
associated increase in density and temperature. The proper density of the
plasma in the precursor is compressed by a factor of up to $\sim 10$. The
corresponding increase in temperature is actually greater than that expected
from adiabatic heating alone. This additional heating comes from the
dissipation of Poynting flux, which already starts at the leading edge of the
precursor. Below, we attribute such a shock structure to the effects of
superluminal electromagnetic waves. We call this novel structure an
electromagnetically modified shock, in close analogy with the so-called
cosmic-ray modified shock \citep[e.g.,][]{1981ApJ...248..344D}. Following the
convention established in that field, the shock-like discontinuity will be
called a subshock.

\ifemulateapj
\FigureTwo
\fi

Note, however, that a substantial amount of Poynting flux dissipation is
evident not only in run~A but also in run~B (the bottom panel of
Fig.~\ref{fig1}). Clearly, the shock-like discontinuity observed in run~B is
not just an ordinary MHD shock, at which the dissipation of Poynting flux
would be strongly suppressed. Fig.~\ref{fig2} shows a closeup view of the
electromagnetic fields around the shock-like discontinuity in run~B. In the
bottom panel, non-MHD electric fields defined as $\tilde{E}_y = E_y - u_x B_z
/ \gamma c, \tilde{E}_z = E_z + u_x B_y / \gamma c$ are shown. Both upstream
and downstream of the discontinuity, $\tilde{E}_{y,z}$ is essentially zero,
indicating that MHD is a fairly good approximation in these regions. Around
the discontinuity, however, non-MHD electric fields have large amplitudes
comparable to MHD type electric fields. These non-MHD electric fields provide
a sort of anomalous resistivity, as is apparent from the generalized Ohm's law
(see Appendix~\ref{numerical}),
\begin{eqnarray}
 && \mb{E} + \frac{\mb{V}}{c} \times \mb{B} =
  \sum_{s} \frac{1}{q_{s} N} \times \eqnewline{&&}
  \left(
   \frac{\del}{\del t} \left( \frac{w_{s}}{c^2} \gamma_{s} \mb{u}_{s} \right) +
   \div \left( \frac{w_{s}}{c^2} \mb{u}_{s} \mb{u}_{s} + p_{s} \mb{I} \right)
  \right), \label{eq:general_Ohm}
\end{eqnarray}
where $N = \sum_{s} \gamma_{s} n_{s}$, $\mb{V} = \sum_{s} n_{s} \mb{u}_{s} /
N$. (This expression is a generalization to warm fluids of the Ohm's law found
by \citet{1996MNRAS.279.1168M}.) A non-zero left-hand side of
equation~(\ref{eq:general_Ohm}) introduces two-fluid effects, which cause
dissipation of the electromagnetic energy and heating of the plasma.

On the downstream side of the shock-like discontinuity, an interface is
formed, which separates magnetized and unmagnetized regions (not visible in
Fig.~\ref{fig1}, since it has already left the region shown, although still
contained within the simulation box). This interface separates the plasma
which was initially downstream, from that which was initially upstream, and
can be called a contact-like interface, in analogy with hydrodynamics.  As in
hydrodynamics, this interface travels with the local fluid speed, which is
non-relativistic. This severely limits the Poynting flux carried by the
downstream flow, so that a quasi-steady state can be achieved only if the
incoming Poynting flux is dissipated in the shock-like discontinuity.
Otherwise, a magnetized MHD shock would form and move rapidly into the
upstream region.  Obviously, the dissipation is made possible because of the
additional degree of freedom provided by the oscillation of the
electromagnetic fields. However, we will not analyze this run in depth here,
but instead focus on the high frequency regime $\Omega/\omega_{p0} > 1$ in
which formation of the novel precursor structure is observed. As we show in
section~\ref{discussion}, this is motivated by the fact that the low density
regime exemplified by run~A is more likely to correspond to a realistic pulsar
wind termination shock than the high density regime of run~B.

\subsection{Time Evolution}
\label{evolution}

Fig.~\ref{fig3} shows the time evolution for run~A. Already at the very
beginning of the simulation when the magnetized upstream plasma starts to
interact with the downstream unmagnetized hot plasma region, the incoming
Poynting flux is dissipated at the subshock. (Here we use the term
``subshock'' for the discontinuity in density rather loosely, since the shock
has not yet developed into a modified one.) As time progresses, the
dissipation front starts to propagate upstream, while the position of the
subshock stays almost constant. The advance of the dissipation front ceases at
around $\omega_{p0} t \sim 1000$, and the system relaxes to a quasi-stationary
state.

\ifemulateapj
\FigureThree
\fi

Since the upstream flow is supersonic, the only way to transfer any
information toward upstream against the flow is through superluminal
waves. Such waves may easily be generated when the magnetic shear waves
transmit through the subshock, because the oscillation frequency measured in
the shock (or downstream) frame is higher than the local proper plasma
frequency. While these waves tend initially to propagate in the positive
$x$-direction, this is actually prevented by the downstream plasma. As can be
seen in Fig.~\ref{fig3}, there is a region behind the subshock in which finite
Poynting flux remains. Beyond that, a region occupied by an unmagnetized
plasma exists (identified by essentially zero Poynting flux). These two
regions in the downstream are separated by a contact-like interface as in the
case of run~B. This interface is characterized by a large density hump, making
it overdense, and prohibiting the penetration of superluminal waves. Here, as
in run~B, the contact-like interface propagates with the non-relativistic
local flow speed. Consequently, superluminal waves are forced to reflect, and
propagate in the negative $x$ direction. This is the reason why backward
superluminal waves are generated, which leak out through the subshock and
interact with the upstream. They affect the upstream plasma in two ways: they
decelerate the flow, and increase its temperature into the relativistic
regime. This forms a precursor region ahead of the subshock.  Reflection from
the contact-like interface is clearly an artifact of the initial
conditions. However, once the shock has evolved into a quasi-stationary state,
the mechanism by which backwards propagating superluminal waves are generated
changes into one related to a parametric instability in the downstream region,
as we discuss below.

It is interesting to note that the initial development resembles the evolution
of a Riemann problem in fluid dynamics, or, in fact, in any set of hyperbolic
partial differential equations. In this analogy, the propagation of the
dissipation front toward upstream corresponds to a rarefaction fan. A
rarefaction fan in hydrodynamics is a smooth transition of an unperturbed
fluid to a rarefied (low pressure) state. In our case, the electromagnetic
energy density in the upstream region decreases across a smooth transition
layer. This similarity is not a mere coincidence. The carrier of information
in our case is a superluminal wave, whose group velocity exceeds the flow
speed. The upstream flow is thus ``subsonic'' with respect to this wave,
resulting in the formation of a rarefaction-like smooth transition. We
conjecture that this transient phase is eventually terminated when the system
relaxes into a state in which the smooth transition of the precursor, followed
by the subshock, satisfies the boundary condition imposed by the downstream
state.

\subsection{Modified Shock Structure}
\label{structure}

\ifemulateapj
\FigureFour
\fi

An enlarged view of the precursor region of run~A is shown in Fig.~\ref{fig4},
at $\omega_{p0}t = 1800$, after the system has reached a quasi-steady
state. The leading edge is located at around $x \sim 1850 c/\omega_{p0}$,
beyond which the dissipation of Poynting flux starts. Although the bulk flow
decelerates substantially in the precursor, its speed remains relativistic.
The subshock can be clearly identified as a discontinuous jump in the flow
speed. The Poynting flux gradually decreases in the precursor and then starts
to fluctuate with large amplitude around the subshock $x \sim 2100
c/\omega_{p0}$.

There is no obvious jump in Poynting flux across the subshock, which satisfies
the Rankine-Hugoniot relations for a relativistic shock in an unmagnetized
plasma; i.e., the electromagnetic fields remain unchanged across it. This is a
clear demonstration of non-MHD behavior, since it implies $\mb{E} \ne -
\mb{V}/c \times \mb{B}$. Thus, the plasma in the precursor is no longer
magnetized, and the waves are essentially superluminal in nature.  This is
confirmed in the second panel from the bottom of Fig.~\ref{fig4}, which
clearly shows $E/B > c$ in the precursor.

To reveal the nature of waves in more detail, we decompose, in
Fig.~\ref{fig4}, the transverse electromagnetic fields into positive and
negative helicities (as defined in section~\ref{model}). Although MHD and
non-MHD type electric fields of the same helicity are not clearly separated by
this procedure (unless the magnitude of $E$ and $B$ are appreciably
different), the generation of waves with a helicity opposite to that of the
injected wave may easily be identified. Hereafter, positive and negative
helicity field components are respectively denoted by $^{+}$ and $^{-}$
superscripts. Recall that the injected magnetic shear wave has a positive
helicity ($R^{+}$) in the simulation.

The second and third panels from the top in Fig.~\ref{fig4} show the negative
and positive components of $E_y$ and $B_z$, respectively.  The amplitude of
the injected positive helicity wave is modulated in the precursor, but shows
an overall decreasing trend. Coherent negative helicity waves are also seen in
the precursor, although they are essentially absent in the upstream. The
anti-correlation in phase of $E_y^{-}$ and $B_z^{-}$ implies that this
component carries a negative Poynting flux, and that the waves are propagating
against the flow.  This can also be seen in the top panel, which shows a
$t$~-~$x$ diagram of the negative helicity component $E_y^{-}$. These waves,
however, disappear when they approach the leading edge of the precursor. This
can be understood from the spatial profile of the proper plasma frequency
$\omega_{p}$ shown in the bottom panel. Because the plasma in the precursor is
heated to relativistic temperatures, $\omega_{p}$ {\it decreases}, despite the
fact that plasma is compressed in the decelerating flow. This effect can occur
only in a relativistically hot plasma, where, because the particle rest-mass
is negligible, one has $w\propto p$ and, therefore, $\omega_{p}\propto
T^{-1/2}$.  Since $\omega_{p}$ coincides with the cutoff frequency of a
circularly polarized superluminal wave, lower frequency waves are permitted in
the regions of relativistically hot plasma, but are excluded from the cooler
upstream plasma. As we show in the next section, the negative helicity waves
have frequency lower than $\omega_{p0}$, the cutoff frequency in the far
upstream region.  (This might also be expected from the longer wavelengths
that are apparent in the second panel from the top of Fig~\ref{fig4}.)  These
waves are, therefore, unable to propagate into the far upstream. It is
difficult to decide whether they are reflected or absorbed when they
encountered the overdense region. Nevertheless, it is clear that these waves
are the only agents that can carry information from the downstream to the
precursor, and trigger the mode conversion from an entropy mode to a
superluminal wave. We conclude that they are essential for the formation of
the modified shock.

\subsection{Wave Spectra}
\label{spectra}

\ifemulateapj
\FigureFive
\fi

The wave properties may be further clarified by looking at their Fourier space
representations. Fig.~\ref{fig5} shows the power spectral density of (in the
left panel) $E_y - i E_z$, (center) $B_y - i B_z$, and (right) $n$ in
$k$~-~$\omega$ space. The spectra are calculated for the time interval $1544
\leq \omega_{p0} t \leq 1800$ using the Blackman window to remove edge effects
and improve the dynamic range. As mentioned earlier, we do not observe
Langmuir-like waves (or charge density fluctuations) in the simulation. The
density fluctuation is thus solely attributable to sound-like waves (positron
and electron densities oscillating in phase). In each panel, the driving
frequency $\Omega/\omega_{p0} = 1.2$, and a theoretical dispersion relation
are shown as dashed lines.  In the left and center panels, the dispersion
relation is that of a circularly polarized superluminal mode
(\ref{dispersionrelation}), with the value of the proper plasma frequency
computed as an average over the appropriate space and time interval. In the
right panel the dispersion relation is that of sound waves moving along $x$
with speed $\pm c_s=\pm\sqrt{\Gamma p/w}$ in a frame moving at speed
$u_x/\gamma$:
\begin{eqnarray}
\omega&=&k\frac{u_x\pm c_s\gamma}{\gamma\pm c_s u_x}
\label{soundwave}
\end{eqnarray}
with the fluid quantities again computed as averages over the appropriate
domain. Note that the forward and backward propagating (with respect to the
comoving frame) sound waves are almost identical when the flow speed is
relativistic, whereas the dispersion relation of the superluminal mode is
independent of the plasma streaming speed, and depends only on the local
proper density and temperature (see Appendix~\ref{superluminal}). Thus, in
the upper right panel (the density panel of row (a)), the two sound waves are
almost indistinguishable. In this panel we plot an additional (dash-dotted)
curve, which we discuss below.

At the leading edge of the precursor (panels (a), $1850 < x / c/\omega_{p0} <
1952$) where the incoming Poynting flux starts to dissipate, a peak of the
field intensities in the first quadrant ($k>0$, $\omega>0$, corresponding to
$R^{+}$) at the driving frequency $\Omega/\omega_{p0} = 1.2$ is clearly
seen. One also finds small amplitude waves around the peak (i.e., the
excitation of sidebands) as well as on the $R$-mode ($\omega > 0$) theoretical
dispersion branch. The density spectrum looks rather
structureless. Nevertheless, it seems to consist of two parts; one on $\omega
= k c$ and another on the dash-dotted curve shown in the right panel of
Fig.~\ref{fig5}(a).

Deep within the precursor (panels (b), $2000 < x / c / \omega_{p0} < 2102$)
where we observe backward propagating negative helicity waves, the power
distribution is very different. The peak at the driving frequency is not so
clear, though the power is still strongest at this point. Instead, the power
is distributed rather broadly. In addition, there is now substantial power in
negative helicity backward propagating waves (second quadrant, $k<0$,
$\omega>0$, corresponding to $R^{-}$) as also seen in Fig.~\ref{fig4}.  The
power of these waves is concentrated in the range $\omega / \omega_{p0} \simeq
0.4 {\rm -} 0.8$ (i.e., below the cutoff frequency in the far upstream),
consistent with their disappearance during propagation in the precursor. The
power of the density spectrum in this region is almost concentrated on $\omega
= k c$.

In the downstream region (panels (c), $2150 < x / c / \omega_{p0} < 2252$),
the power distribution is also broad and lower frequency waves have relatively
large amplitudes whereas the peak structure at the driving frequency
persists. Since low frequency waves with negative helicity (second quadrant,
$R^{-}$) have upstream directed group velocities, they can propagate to the
subshock and leak out into the precursor where the cutoff frequency is even
lower. The top panel of Fig.~\ref{fig4} supports this interpretation. One can
see there that the backward propagating waves are continuously propagating
from the downstream to the precursor with refraction at the subshock,
consistent with the idea of leakage. We think this is the reason why there are
backward propagating superluminal waves with lower frequencies in the
precursor. In the density spectrum, the forward and backward propagating sound
waves are now clearly separated because the flow speed is
non-relativistic. However, the forward branch carries much more power than the
backward one.

The evolution of power spectrum described above may be explained qualitatively
in terms of parametric instabilities or nonlinear wave-wave
interactions. Fig.~\ref{fig6} depicts possible nonlinear couplings of a pump
superluminal wave with different wave modes that are likely to be taking place
(a) in the precursor and (b) downstream. Note that the figure shows the
coupling between waves as measured in the simulation frame. It is known that
nonlinear interactions among waves become strongest when the matching
condition for both frequency and wavenumber is satisfied:
\begin{eqnarray}
 \omega_{\pm} &=& \omega_{0} \pm \omega \\
 k_{\pm} &=& k_0 \pm k
\end{eqnarray}
Here $(\omega_0, k_0)$, $(\omega, k)$, $(\omega_{\pm}, k_{\pm})$ describe,
respectively, the pump wave, a sound-like daughter wave, and
electromagnetic-like daughter waves. Here we consider only a sound-like wave
as a longitudinal perturbation, so the process may be called stimulated
Brillouin scattering. It is generally expected that the growth rate of an
instability is largest when the generated daughter waves lie close to the
normal modes of the unperturbed plasma. Although, in principle, coupling to a
daughter wave far from the normal modes may occur (such as in the case of
$(\omega_{+}, k_{+})$ in panel (b) of Fig.~\ref{fig6}), the power is usually
expected to be much smaller.

\ifemulateapj
\FigureSix
\fi

In the density spectrum measured in the precursor, sound mode waves almost
coincide with $\omega = k c$. Thus, for a superluminal pump wave with
$\omega/\omega_{p} \gg 1$ in the laboratory frame, all relevant waves (i.e.,
electromagnetic and sound waves) are almost aligned with the straight line
$\omega = k c$ in $k$~-~$\omega$ space. In this case, there are many possible
wave-wave couplings that satisfy the matching condition, as can be envisaged
from Fig.~\ref{fig6}(a). Notice that once sideband waves are generated, they
may subsequently decay via exactly the same process, i.e., the instability
occurs recursively, generating many sidebands, and eventually leading to a
turbulent spectrum. The generation of sideband modes results in amplitude
modulation (i.e., a beat wave) in real space. This characteristic feature is
clearly seen in both Figs.~\ref{fig4} and \ref{fig5}. In Fig.~\ref{fig4}
(third panel from the top), the positive helicity component in the precursor
exhibits amplitude modulation. The sideband modes causing this modulation are
seen in the power spectrum measured at the same region Fig.~\ref{fig5}(b).

Note that sideband generation, albeit of small amplitude, has already started
in Fig.~\ref{fig5}(a), supporting the idea that the incoming wave has already
been converted into a superluminal wave at the leading edge of the
precursor. The power in density perturbations in this region is concentrated
at low frequencies, consistent with the above picture (see
Fig.~\ref{fig6}(a)). However, substantial power can also be seen lying roughly
on the dash-dotted curve in the density spectrum in Fig.~\ref{fig5}(a). This
may be explained as follows: if one considers not only a superluminal mode
wave but also the incoming shear wave as the pump waves (i.e., finite
amplitude), the coupling of these two waves may produce density
fluctuations. Since superluminal waves of finite amplitude are found on the
theoretical $R$ mode dispersion branch, the matching condition between the
superluminal waves and the incoming shear wave implies that the frequency and
wavenumber $(\omega, k)$ of the resulting density fluctuation must obey
\begin{eqnarray}
\omega &=& \Omega - \sqrt{\omega_p^2 + (\Omega_0 - k V_0)^2
  c^2/V_0^2}
\label{densityfluctuations}
\end{eqnarray}
which is plotted as the dot-dashed curve in Fig~\ref{fig5}. Such a dispersion
relation is far from normal modes of dispersion relation of the unperturbed
plasma. It can exist only when the wave is forced to oscillate because of the
coupling between multiple finite amplitude waves. Thus, the presence of
enhanced fluctuations around this curve implies that coupling between the
injected magnetic shear wave and superluminal modes is indeed taking place,
and could be the mechanism responsible for mode conversion. This
interpretation is supported by the absence of such density fluctuations deep
inside the precursor (b), confirming that the shear wave has been fully
converted into superluminal modes at this point.

In the downstream region the forward and backward sound waves are clearly
separated (see the density power spectrum in Fig.~\ref{fig5}(c)), and it is
more natural to invoke the coupling depicted in Fig.~\ref{fig6}(b) producing a
backward propagating electromagnetic-like wave. The backward propagating low
frequency superluminal waves thus may be interpreted as a result of the back
scattering of the pump wave off a sound-like mode wave, i.e., backward
Brillouin scattering.  This interpretation is consistent with the enhanced
forward propagating sound-like wave activity seen in Fig.~\ref{fig5}(c). These
waves are seen to steepen, producing many small-scale shocks propagating in
the forward direction, which, ultimately, dissipate energy due to the
numerical viscosity inherent in fluid simulations.  As mentioned above, the
backward propagating superluminal waves propagate into the precursor and
trigger the conversion of the incoming wave mode at its leading edge, thus
causing the modification of the shock structure.

\subsection{Dissipation Efficiency}
\label{dissipation}

Although we have discussed some elementary processes leading to the
dissipation of electromagnetic energy, it is difficult to estimate the actual
dissipation efficiency and its parameter dependence from an analytical
treatment alone, because this involves complicated nonlinear processes. Other
simulations we have performed with different $\gamma_0$ and $\sigma_0$ for a
fixed $\Omega/\omega_{p0}$ show that the basic processes seem to be
essentially the same, and the shock structure is always strongly
modified. Fig.~\ref{fig7}, for example, shows the spatial profiles of flow
velocity $u_x$ and $\sigma$ for three runs with different initial Lorentz
factors: $\gamma_0 = 20$, $40$ (run~A), and $60$ while $\sigma_0=10$ and
$\Omega/\omega_{p0}=1.2$ are kept constant. In these runs (and also in others
we have performed but do not show here) $\sigma$ is reduced to below unity in
a precursor region, and appears unaffected by the subshock transition.  In the
downstream region, it typically goes down to $\sim 0.1$, although it exhibits
strong fluctuations.  However, the incoming wave is converted into
superluminal waves, implying that the remaining Poynting flux is entirely
carried by these waves, and the downstream plasma is essentially unmagnetized.
Therefore, the conventional MHD picture does not apply, even though a finite
electromagnetic field remains. Furthermore, superluminal waves in the
downstream continue to dissipate due to instabilities.  Thus, the $\sigma$
values obtained in simulations may give only an upper limit.

\ifemulateapj
\FigureSeven
\fi

\section{Discussion}
\label{discussion}

Dissipation in Poynting-flux dominated relativistic flows is an important
issue in high energy astrophysics. Although the possibility that superluminal
waves may play a role has been discussed for decades
\citep[e.g.,][]{1976JPlPh..15..335K,1978A&A....65..401A,1980A&A....83....1K,
1996MNRAS.279.1168M},
it has not been known how such waves can be generated in a self-consistent
manner. In this paper we have analyzed in depth a specific example that shows
how strong superluminal waves can be generated by a relativistic shock when it
interacts with an inhomogeneous upstream plasma.  We selected a situation
where the inhomogeneity takes the form of a periodic magnetic shear embedded
in a cold plasma. The frequency $\Omega$ of the shear wave measured in the
shock frame was chosen to be greater than the proper plasma frequency
$\omega_{p}$ in the upstream. The superluminal wave was observed to be
generated as a result of mode conversion from the injected shear wave.

Once the conversion has taken place, superluminal waves may become unstable to
various types of instabilities
\citep[e.g.,][]{1973PhFl...16.1480M,1978JPlPh..20..313L,1978A&A....65..401A}.
In the present paper, we have shown that the decay process proceeds via
stimulated Brillouin scattering. Sound-like waves grow to large amplitude via
this instability. They then steepen to form small-scale shocks that cause
strong heating of the plasma. The heating results in a pressure gradient that
decelerates the flow, forming a precursor ahead of the subshock. This process
requires a trigger, because otherwise the upstream medium is uniform. We
suggest that this role is played by backward propagating superluminal
waves. Although their amplitude in the laboratory frame is relatively small
compared to the incoming wave, it is easily shown that the situation is the
opposite when viewed in the rest frame of the incoming plasma. Therefore, we
think they are responsible for the instability that triggers the formation of
precursor.  Although this is the principal finding of this paper, our
understanding of the formation mechanism remains incomplete.  Important
questions, such as the precursor scale length, the compression ratio, and the
stability of the shock, remain open. Our simulations also leave open the
question of \lq\lq spontaneous\rq\rq\ conversion (i.e., well upstream of the
termination shock) of the subluminal mode into a superluminal mode.

It is well-known that a strong electromagnetic wave can also suffer Raman
scattering off a Langmuir-like wave, as well as Brillouin scattering off a
sound-like wave.  However, our simulations show no evidence of charge density
fluctuations. We attribute this to the symmetry between electron and positron
motions in the presence of transverse waves. It is easy to check that the
transverse velocities of two fluids have the same magnitude (and opposite
directions) for both the magnetic shear and superluminal mode. A first order
longitudinal velocity perturbation is affected by the presence of a pump
transverse wave through the Lorentz force $q \mb{u} \times \mb{B} /\gamma c$,
so that the terms are exactly the same for the electrons and positrons, both
in magnitude and direction, when the above mentioned symmetry holds. This
leads to sound-like perturbations, rather than charge-density perturbations.
From this consideration, it appears natural that we do not observe the growth
of Langmuir waves. This symmetry will be broken in the presence of a finite
longitudinal magnetic field component. However, a detailed analysis of
possible parametric instabilities is left for future work.

In the present paper, we analyzed the case $\Omega/\omega_{p0} =1.2$,
$\sigma_0=10$, where superluminal modes of frequency $\Omega$ could, in
principle, propagate in both in the upstream and downstream regions, if these
regions were free of other electromagnetic fields.  We found that a hot
precursor region is formed, in which the plasma frequency is reduced, and
which contains waves whose frequency lies below the cut off in the upstream.
This did not happen in the comparison case (run~B) in which
$\Omega/\omega_{p0}=0.4$, $\sigma_0=10$, because the downstream plasma
frequency is higher than the wave frequency. Although the Poynting flux was
dissipated in the structure we found in run~B, there was no evidence of
superluminal waves, nor of shock modification. In the absence of a survey of
parameter space, we cannot reach a definite conclusion on the range of
frequencies and magnetizations that will produce an electromagnetically
modified shock structure.  However, our interpretation suggests that the
ability of the {\em downstream} shocked plasma to support superluminal waves
is crucial, and that their propagation in the unperturbed upstream medium
(which in any case contains other wave fields of large amplitude) is not
essential.

Our approach differs in concept from that of \citet{2005ApJ...634..542S}, who
performed particle-in-cell simulations of a shock front interacting with an
injected superluminal wave of prescribed frequency and amplitude.
Nevertheless, the results are superficially similar, in the sense that
large-amplitude, forward propagating, superluminal waves exist upstream of the
shock and were found to decrease in amplitude as they approach it.  The role
of backward-propagating modes, and that of density fluctuations and stimulated
Brillouin scattering are not evident in their results.  This may be because of
the different driving mechanisms studied, or because the additional effects
captured in a 2.5 dimensional particle-in-cell simulation intervene to reduce
their importance. On the other hand, it may simply be due to the difficulty of
resolving these features in an approach that is intrinsically more
computationally intensive. Further work is clearly needed to resolve these
issues.

In the case of a pulsar wind with a termination shock whose position is
approximately stationary with respect to the central object, it is fairly easy
to estimate the condition for the possible existence of superluminal
waves. Assuming a radial wind in which the density falls off as $1/r^2$ ($r$
is the distance from the central object), the condition for superluminal waves
to propagate in the shocked wind, $\Omega\sqrt{\sigma}/\omega_{p}>1$, may be
written as \citep{2012ApJ...745..108A}
\begin{eqnarray}
\frac{r}{r_{\rm L}}&>& 2.8\times10^6
 \left( \frac{L}{10^{38} \, {\rm erg/s}} \right)^{-1/2}
 \left( \frac{\dot{N}}{10^{40} \, {\rm s^{-1}}} \right)
\label{estimatercrit}
\end{eqnarray}
where $r_{\rm L}$ is the radius of the light cylinder, $L$ is the wind
luminosity (assuming spherical symmetry) and $\dot{N}$ is the flux of electron
positron pairs carried by the wind.  Estimates of $\dot{N}$ vary widely from
pulsar to pulsar. From observations of the Crab Nebula (whose pulsar has
$L=4\times10^{38}\,\textrm{erg/s}$), one finds
$\dot{N}\approx10^{40}\,\textrm{s}^{-1}$, assuming this rate has been more or
less constant over the history of the nebula. Thus, in the case of the Crab,
where the termination shock identified on X-ray images
\citep{2000ApJ...536L..81W} is estimated to lie at roughly $10^9 r_{\rm L}$,
we expect this structure to be strongly modified by electromagnetic
waves. Only in the cases where the pulsar is surrounded by a high pressure
medium, such as the wind of a close companion star, can one expect the
termination shock to lie sufficiently close to the pulsar to be described by
an MHD model \citep{mocholkirk12}. In the case of the double pulsar system
J0737-3039, for example, the lack of orbital modulation of the observed X-ray
emission \cite{2007ApJ...670.1301C} places a constraint on the radiative
efficiency of the termination shocks, which has potential implications for
their structure. However, the absence of a reliable estimate of $\dot{N}$ for
these pulsars renders the interpretation difficult.

Particles moving in a superluminal wave emit radiation that is usually
called either synchro-Compton or nonlinear inverse Compton radiation.
We have ignored such radiation losses in our simulations. This can
be justified by making a rough estimate of the ratio of the
cooling time of a single electron and the dynamical timescale of interest.
Taking for the latter the inverse of the wave frequency, we find:
\begin{eqnarray}
\Omega t_{\rm cool}&\approx& \left(\frac{2 e^2\Omega}{3 m c^3}\right)^{-1}
a^{-3}
\end{eqnarray}
where the strength parameter of the wave is
\citep[e.g.,][]{2011ApJ...729..104K}
\begin{eqnarray}
a&=& 3.4\times10^{10}
 \left( \frac{r_{\rm L}}{r} \right)
\left( \frac{L}{10^{38} \, {\rm erg/s}} \right)^{1/2}
\end{eqnarray}
Thus, for a given pulsar (fixed $\Omega$ and $L$), $\Omega t_{\rm cool}\propto
r^3$. For example, in the case of the Crab, we find
\begin{eqnarray}
\Omega t_{\rm cool}&\approx& 2.4\times10^{-12}\left(r/r_{\rm L}\right)^3
\label{radlosses}
\end{eqnarray}
confirming the finding by \citet{1978A&A....65..401A} that radiation damping
is important for waves that propagate close to the light-cylinder of this
pulsar, but not at the termination shock.  However, for realistic pair loss
rates, superluminal waves cannot propagate close to the star. Comparing
(\ref{radlosses}) with (\ref{estimatercrit}) shows that $\Omega t_{\rm
cool}>1$ everywhere in the propagation zone of a pulsar wind provided that
\begin{eqnarray}
\dot{N}&>&4\times 10^{36}
\left( \frac{L}{10^{38} \, {\rm erg/s}} \right)
\left( \frac{P_{\rm pulsar}}{1 \, {\rm s}} \right)^{-1/3}
\end{eqnarray}
where $P_{\rm pulsar}$ is the pulsar period. Therefore, radiation losses can
be neglected at the termination shock unless this lies close to the critical
radius, and the pulsar has a rather large mass-loading parameter:
$\mu=L/\left(\dot{N}mc^2\right)>3\times 10^7
\left(P_{\rm pulsar}/1\,\textrm{s}\right)^{1/3}$.

Superluminal waves may also be important in other high energy astrophysical
objects. In the present paper, we consider only a coherent upstream wave,
because of its relevance to pulsar winds and also for simplicity. However, any
magnetic field irregularities in the upstream medium may be converted into
superluminal modes by a shock when their spatial scales are small enough so
that the corresponding ``frequency'' measured in the shock frame is greater
then the cutoff frequency in the shocked plasma. For more general
applicability, the effects of a phase-averaged magnetic field component and of
different polarizations needs to be investigated in more detail. Naively, we
expect that it is only the oscillating magnetic field components that can be
dissipated by this process. A finite, phase-averaged magnetic field, such as
expected in the off-equator regions of pulsar winds, will, therefore, remain
undamped, so that the prominent jet-torus structure that is explained by MHD
models would be reproduced \citep{2004MNRAS.349..779K,2012arXiv1212.1382P}.

Our model obviously omits kinetic effects. This may at least be partially
justified when the plasma is strongly magnetized, $\sigma \gg 1$, so that the
dispersion effects inherent in the two-fluid equations set in earlier than
kinetic effects such as that of a finite Larmor radius. Moreover, as far as
superluminal waves are concerned, the two-fluid approximation is adequate,
because, by definition, they cannot suffer collisionless damping through
resonant wave-particle interactions. On the other hand, if the superluminal
waves generate longitudinal oscillations, kinetic effects could play an
essential role. In the two-fluid approximation, a linear sound wave is an
eigenmode of the system. On the other hand, it is well known that an ion
acoustic wave in a non-relativistic electron-ion plasma with the same
temperature is heavily damped. The reason is that the phase speed of the ion
acoustic wave is of the same order of the ion thermal velocity, so that
thermal ions can absorb energy from a wave via Landau resonance. This damping
is even more severe in a relativistic pair plasma because the phase velocity
of a sound wave is limited by $c/\sqrt{3}$, whereas individual particles
travel essentially with $c$. Therefore, the dissipation may, in reality, occur
through immediate absorption of sound-like waves by relativistic particles,
without producing small-scale shocks. Alternatively, in situations where Raman
scattering dominates over Brillouin scattering, the dissipation may be caused
by nonlinear collapse of Langmuir waves.

Another important property of electromagnetically modified shocks is that the
subshock behaves as if it were unmagnetized.  This may be a particularly
promising scenario for particle acceleration.  Conventionally, acceleration at
a strongly magnetized relativistic shock is considered to be much more
difficult than at a non-relativistic shock, because the former are generically
superluminal
\citep{1990ApJ...353...66B,2004ApJ...610..851N,2009ApJ...698.1523S}. In such a
shock, the only way for a downstream particle to cross the shock front is by
cross-field diffusion, which cannot compete with advection by a plasma moving
at a significant fraction of the speed of light.  In the present scenario,
however, we conjecture that particles would more easily cross the subshock
because they are no longer magnetized
\citep{2007ApJ...668..974K,2008ApJ...682L...5S}, and, therefore, do not have
to rely on cross-field diffusion.  In this case, the electromagnetically
modified shock would be an efficient particle accelerator.

Finally, we mention the striking similarity between electromagnetically
modified and cosmic-ray modified shocks. In our case, forward superluminal
waves, generated in association with the precursor, eventually produce back
scattered waves in the downstream. These return to the precursor, and then
operate as a catalytic agent causing the mode conversion: This is how the
quasi-stationary modified shock structure is maintained. In a cosmic-ray
modified shock, cosmic rays play precisely the same role. They diffuse around
the shock and thus can easily leak out from the downstream to the upstream due
to their large thermal velocities. There, they excite Alfvenic waves that
exert a pressure on the background plasma, leading to deceleration of the flow
in the precursor. The injection of cosmic rays is thought to occur primarily
at the subshock, whereas superluminal waves are created at the leading edge of
the precursor in the case of an electromagnetically modified shock. It would be
interesting to investigate how these two different shock modification concepts
might be combined with each other, and what effect this would have on particle
acceleration.

\section{Summary}
\label{summary}

We have investigated the interaction between a relativistic shock and a
circularly polarized sinusoidal magnetic shear wave embedded in an upstream
strongly magnetized plasma, mimicking a pulsar wind termination shock. We
found that large amplitude superluminal electromagnetic waves are generated
through mode conversion, and subsequently strongly modify the shock structure,
producing a precursor region ahead of a subshock. In association with the
formation of the modified shock structure, a substantial amount of Poynting
flux is dissipated. The dissipation mechanism is interpreted as a stimulated
Brillouin scattering process. The superluminal waves suffer a parametric
instability and generate sound-like waves.  These waves then steepen into
shocks which subsequently heat the plasma. The process is considered to be an
efficient mechanism for converting the dominant electromagnetic energy into
particle kinetic energy. The remaining Poynting flux in the downstream is
carried by superluminal waves rather than by a magnetized MHD flow. Moreover,
dissipation can be expected to continue until the oscillating fields die away
altogether. Therefore, the downstream state will essentially be an
unmagnetized relativistic hot plasma. A simple estimate for the applicability
of this model is given, suggesting that this process will play a crucial role
at least at the termination shocks of young, rapidly rotating, isolated
pulsars.

\acknowledgments T.~A. is supported by the Global COE program of
Nagoya University (QFPU) and KAKENHI 22740118 from JSPS of Japan. Most
of this work was carried out while one of the authors (T.~A.) was
visiting Max-Planck-Institut f\"ur Kernphysik through the
Institutional Program for Young Researcher Overseas Visits of Nagoya
University from JSPS of Japan.

\appendix

\section{Circularly Polarized Superluminal Wave}
\label{superluminal}

An analytic solution of a circularly polarized superluminal
electromagnetic wave in a finite (relativistic) temperature plasma was
first given by \cite{1973PhFl...16.1277M}. Since a circularly
polarized mode does not involve any perturbations in density,
pressure, or longitudinal velocity, the finite temperature effect
appears only as a factor increasing the inertia of the fluids. We
consider here the case without phase-averaged magnetic fields and, thus,
the electromagnetic fields are those of waves only. Assuming a magnetic
field perturbation of arbitrary amplitude in the form
\begin{eqnarray}
 B_y &=& B_0 \cos (k x - \omega t) \\
 B_z &=&-B_0 \sin (k x - \omega t),
\end{eqnarray}
the electric field is calculated by Faraday's law as
\begin{eqnarray}
 E_y &=&-\frac{\omega}{k c} B_0 \sin (k x - \omega t) \\
 E_z &=&-\frac{\omega}{k c} B_0 \cos (k x - \omega t).
\end{eqnarray}
Then, it follows from the positron's equation of motion
\begin{eqnarray}
 u_y &=& \frac{e}{k h m c} B_0 \cos (k x - \omega t) \\
 u_z &=&-\frac{e}{k h m c} B_0 \sin (k x - \omega t),
\end{eqnarray}
where a factor $h = (1 + \Gamma/(\Gamma-1) T/m c^2)$ introduced in the
text is the effective increase of fluid inertia due to relativistic
temperature. Note that the electron's transverse velocity may be
obtained simply by changing sign of the charge. It is easy to verify
that the above relations exactly satisfy Ampere's law if the
dispersion relation
\begin{eqnarray}
 \omega^2 = \omega_{p}^2 + k^2 c^2
\label{dispersionrelation}
\end{eqnarray}
is satisfied. Here, the effective proper plasma frequency is defined by
Eq.~(\ref{eq:cutoff}). Although this is formally the same as that of a linear
electromagnetic wave in a non-relativistic unmagnetized plasma, it is
important to realize that the cutoff frequency is now given by the proper
plasma frequency and, therefore, depends on the wave amplitude. Note that the
dispersion relation is independent of the reference frame because $\omega_{p}$
is defined using only proper-frame quantities. Thus, in any frame, the
existence condition of such a superluminal wave is always given by $\omega >
\omega_{p}$ where $\omega$ is a laboratory frame frequency.

\section{Numerical Method}
\label{numerical}

By taking sum and difference of the two-fluid equations
(cf.~\cite{1996MNRAS.279.1168M}), the governing equations can be rewritten as
\begin{eqnarray}
 \frac{\del}{\del t} \mb{U} + \div \mb{F} = \mb{S},
\end{eqnarray}
where
\begin{eqnarray}
 \mb{U} =
  \left(
   \begin{array}{c}
    \gamma_p n_p + \gamma_e n_e \\
    (w_p \gamma_p \mb{u}_p + w_e \gamma_e \mb{u}_e)/c^2 + \mb{S}_{\rm EM} \\
    w_p \gamma_p^2 + w_e \gamma_e^2 - (p_p + p_e) + E_{\rm EM} \\
    \gamma_p n_p - \gamma_e n_e \\
    (w_p \gamma_p \mb{u}_p - w_e \gamma_e \mb{u}_e) / c^2 \\
    w_p \gamma_p^2 - w_e \gamma_e^2 - (p_p - p_e)
   \end{array}
  \right),
\end{eqnarray}
\begin{eqnarray}
  \mb{F} =
  \left(
   \begin{array}{c}
    n_p \mb{u}_p + n_e \mb{u}_e \\
    (w_p \mb{u}_p \mb{u}_p + w_e \mb{u}_e \mb{u}_e)/c^2 + (p_p + p_e) \mb{I} -
     \mb{T}_{\rm EM} \\
    w_p \gamma_p \mb{u}_p + w_e \gamma_e \mb{u}_e + \mb{S}_{\rm EM} \\
    n_p \mb{u}_p - n_e \mb{u}_e \\
    (w_p \mb{u}_p \mb{u}_p - w_e \mb{u}_e \mb{u}_e)/c^2 + (p_p - p_e) \mb{I} \\
    w_p \gamma_p \mb{u}_p - w_e \gamma_e \mb{u}_e
   \end{array}
  \right),
\end{eqnarray}
and $\mb{S}_{\rm EM} = c (\mb{E} \times \mb{B}) / 4 \pi$, $E_{\rm EM} =
(\mb{E}^2 + \mb{B}^2) / 8 \pi$, $\mb{T}_{\rm EM} = (\mb{E}\mb{E} +
\mb{B}\mb{B})/4\pi - (\mb{E}^2+\mb{B}^2)\mb{I} / 8 \pi$. Defining the average
laboratory density $N = \sum_{s} \gamma_s n_s$ and three fluid velocity
$\mb{V} = \sum_{s} n_s \mb{u}_s / N$, the right-hand side may be written as
\begin{eqnarray}
  \mb{S} =
  \left(
   \begin{array}{c}
    0 \\
    0 \\
    0 \\
    0 \\
    e N (\mb{E} + \mb{V}/c \times \mb{B}) \\
    e N \mb{V} \cdot \mb{E} \\
   \end{array}
  \right).
\end{eqnarray}
It is easy to understand that the first three equations, when supplemented by
$\mb{E} = - \mb{V}/c \times \mb{B}$ and the magnetic field induction equation,
give the standard relativistic MHD equations when appropriate averages of the
two fluids are employed. The remaining three equations, on the other hand,
describe two-fluid effects. If the initial condition is such that the
difference between the two fluids is small (i.e., MHD is an adequate
approximation), two-fluid effects will remain unimportant unless the frozen-in
condition is violated, in which case the right-hand side of the above
equations become non-zero. In particular, the fifth equation (the difference
between momentum equations) can be identified as the generalized Ohm's law
Eq.~(\ref{eq:general_Ohm}).

By using the above form of the basic equations with a conservative scheme, the
conservation of total density, momentum, and energy are satisfied up to
machine precision. On the other hand, when the two fluid equations are written
separately, the Lorentz force terms appear on the right-hand side as source
terms, and conservation of total momentum and energy are not numerically
guaranteed.  This conservation property is very important especially for the
purpose of our discussion about energy conversion mechanisms. Therefore, we
solve the above equations in our numerical simulation code.

The numerical scheme used in this work is based on the central scheme in
semi-discrete form \citep[e.g.,][]{Kurganov2000a,Kurganov2000b} with fifth
order WENO (weighted essentially non-oscillatory) reconstruction
\citep{Jiang1996}. We solve the all governing equations (including Maxwell's
equations) simultaneously without splitting. The characteristic speed of the
system is thus always the speed of light $c$. The equations are discretized
only in space, so that they can be written as a system of ordinary
differential equations in time. We integrate this system of equations using
the third order TVD-RK (Total Violation Diminishing Runge-Kutta) method
\citep{Shu1988}. By using the WENO methodology, the scheme automatically
introduces numerical diffusion when a discontinuity is detected, which enables
us to resolve ordinary discontinuities with only a few grid points. However,
it is found that the simulations tend to be unstable in cases where strong
instabilities are developing. We thus also add explicit numerical viscosity
with a coefficient that depends on the local second order derivative of the
longitudinal component of four velocity ($u_x$), in order to avoid complete
collapse of simulations.

In solving fluid equations in a conservative form, one needs to calculate
primitive variables ($n$, $\gamma$, $\mb{u}$, $p$) from the updated
conservative variables ($\mb{U}$). In our case, we solve a quartic equation to
obtain $|\mb{u}|$, from which other primitive variables are calculated. It is
possible to verify that there is always a solution for $|\mb{u}|$ that is
positive and real under physical conditions, which is calculated algebraically
by the method described in \cite{2009ApJ...696.1385Z} (see their appendices
for detail).

\bibliographystyle{apj}
\bibliography{reference}
%\input{ms.bbl}

%%% figures for manuscript style
\ifemulateapj
\else
\FigureOne
\FigureTwo
\FigureThree
\FigureFour
\FigureFive
\FigureSix
\FigureSeven
\fi

\end{document}